# Multiple Linear Regression-Enhanced RGB-based Luminescence Thermometry for High-sensitivity Thermal Readout

**Y. Abe[1], M. Szymczak[1], Miguel A. Hernández-Rodríguez[2], M. Runowski[3], L. Marciniak[1,*]**

[1] *Institute of Low Temperature and Structure Research, Polish Academy of Sciences, Okólna 2, 50-422 Wrocław, Poland*

[2]*Departamento de Física, and IUdEA, Universidad de La Laguna Apdo. Correos 456, E-38200 San Cristóbal de La Laguna, Santa Cruz de Tenerife, Spain*

[3]*Faculty of Chemistry, Adam Mickiewicz University, Uniwersytetu Poznańskiego 8, 61-614 Poznań, Poland*

*corresponding author: l.marciniak@intibs.pl

## Abstract

The practical implementation of phosphors for luminescence-based thermal sensing and imaging requires simple, user-friendly approaches that enable convenient and sensitive temperature readout. To the best of our knowledge, this is the first demonstration of combining *RGB*-based thermal imaging using a conventional digital camera with multiple linear regression (MLR) to achieve straightforward and highly sensitive temperature determination and spatially resolved thermal imaging. Importantly, the implementation of the MLR approach enhances the relative sensitivity by more than 3-fold, compared with conventional analysis based solely on the *G/R* or *B/R* intensity ratios. The application of $Ca_3Al_2O_6$:$Mn^{2+}$-$Ce^{3+}$ as a luminescent temperature probe, in which the intensity ratio between the $Ce^{3+}$ and $Mn^{2+}$ emission bands exhibits a pronounced temperature dependence, enables temperature readout through several complementary approaches. These include conventional luminescence intensity ratio thermometry ($S_R = 1.46\%\ K^{-1}$), analysis of the CIE 1931 chromaticity coordinates ($S_{Rx} = 0.45\%\ K^{-1}$ and $S_{Ry} = 0.14\%\ K^{-1}$), as well as *RGB*-based thermal sensing and imaging using a digital camera. This multimodal optical response, together with the accessibility of camera-based readout and the substantial sensitivity enhancement enabled by MLR, establishes a practical strategy for spatially resolved luminescence thermometry.

## INTRODUCTION

The use of thermally induced changes in the spectroscopic properties of phosphors for temperature readout has been well established and applied in numerous fields for many years (*1–6*). In the vast majority of cases, such luminescent thermometers rely on either a ratiometric approach or the kinetics of luminescence(*1–6*). Although these are not the only parameters exhibiting thermal sensitivity, their principal advantage lies in their low or negligible sensitivity (under certain specific conditions) to changes in the local environment and measurement conditions(*7*). Both approaches provide reliable and accurate remote temperature readout. However, in each case, temperature measurement requires the use of complex and often expensive detection systems, such as I) emission spectrometers coupled to the cooled CCD or CMOS cameras; or II) time-resolved spectroscopy setups composed of pulsed lasers and photomultipliers. The cost of such systems, their lack of portability, and the complexity of their operation can limit the accessibility of luminescence thermometry for widespread use. Moreover, these approaches typically allow only for point-wise temperature measurement, meaning that thermal imaging of a surface coated with a luminescent thermometer requires point-by-point acquisition of luminescence spectra or luminescence kinetics(*4*, *8–11*). This is not only time-consuming but, in the case of rapidly changing thermal gradients, also results in the loss of valuable information regarding the spatial distribution of thermal variations.

To address these limitations, an alternative approach to thermal imaging has recently been proposed, employing a conventional digital camera as the detector(*12–17*). In its original implementation, this approach enables temperature readout by capturing two luminescence images using appropriate external optical filters, whose spectral characteristics match the shape of the phosphor's luminescence spectrum(*12–14*, *18*). This solution allows for two-dimensional imaging of thermal changes; however, the procedure of optical filter exchange introduces additional practical difficulties and limits the dynamics of the thermal changes that could be recorded. Consequently, a considerably simpler solution was recently proposed, based on the intensity ratio of luminescence maps recorded in the RGB channels of a digital camera(*15–17*, *19–22*). These signals are automatically captured by any conventional camera, and their ratio can serve as a temperature-dependent parameter. The effectiveness and simplicity of this approach have been demonstrated in previous studies.

When considering a suitable phosphor for this type of thermal imaging, it is essential to ensure that the emission bands of the optically active dopants are selectively located within the spectral ranges corresponding to the sensitivity ranges of the individual RGB channels. In this context, systems co-doped with $Ce^{3+}$ and $Mn^{2+}$ ions appear particularly promising, as $Ce^{3+}$ is responsible for blue luminescence(*23–25*), while $Mn^{2+}$ ions located in an octahedral crystallographic environment give rise to intense red emission. Notably, what distinguishes $Mn^{2+}$ ions from other red-emitting activators is that an increase in temperature induces thermalization of higher-lying vibronic components of the excited state, resulting in a blue shift of the emission band(*26–30*). This effect is particularly significant, as it enables the recording of thermally activated luminescence in the green, G channel, which initially (at low temperatures) detects no signal. This, in turn, allows for high sensitivity of the measurement.

In this work, we demonstrate the possibility of enhancing the relative sensitivity of such thermal imaging through the application of multiple linear regression (MLR) algorithms(*31–35*) (Figure 1). To the best of our knowledge, this constitutes the first demonstration of this approach in filter-free thermal imaging, and as will be shown, it enables a significant increase in thermal sensitivity, which is crucial for reliable thermal imaging. The advantage of MLR in this context lies in its numerical simplicity, which eliminates the need for complex computational algorithms to perform the measurement. To demonstrate this approach, studies were carried out on $Ca_3Al_2O_6:Ce^{3+}$, $Mn^{2+}$ as a function of dopant ion concentration and temperature. Additionally, as it will be shown $Ca_3Al_2O_6:Ce^{3+}$, $Mn^{2+}$ offers other mode of temperature readout confirming its potential for multimodal readout.

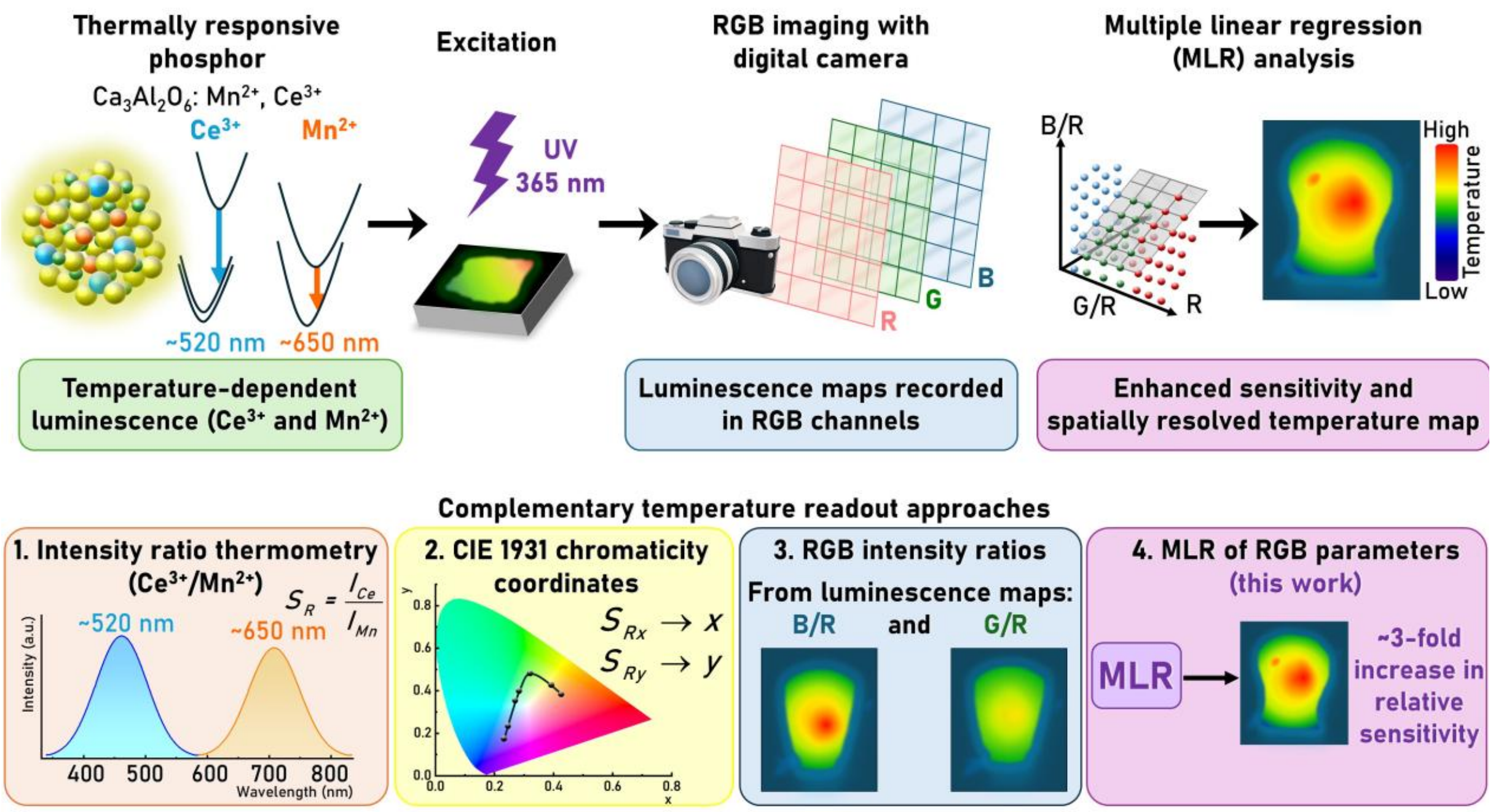

**Figure 1**. Schematic representation of the concept proposed in this work. $Ca_3Al_2O_6$ co-doped with $Ce^{3+}$ and $Mn^{2+}$ is employed as a luminescent thermometer exhibiting a temperature-dependent change in emission color resulting from the different thermal responses of $Ce^{3+}$ and $Mn^{2+}$ luminescence. The emission can be directly recorded using a conventional digital camera, enabling rapid and cost-effective temperature readout based on the temperature-dependent response of the RGB channels. Implementation of multiple linear regression (MLR) analysis using the RGB signals significantly improves the thermometric performance, providing an approximately threefold enhancement of the relative thermal sensitivity compared with the conventional RGB-based approach. The distinct thermal evolution of the $Ce^{3+}$ and $Mn^{2+}$ emission contributions enables multimodal temperature sensing using four complementary readout strategies: (1) spectroscopic ratiometric thermometry, (2) chromaticity-coordinate analysis, (3) RGB-based thermometry, and (4) MLR-assisted RGB thermometry.

## RESULTS

### Structural and morphological characterization

The powder XRD patterns of the $Ca_3Al_2O_6$:$Mn^{2+}$-$Ce^{3+}$ phosphors, together with the reference data of $Ca_3Al_2O_6$ (ICSD 1841), presented in Figure 2a confirm that pure cubic phase of $Ca_3Al_2O_6$ with the *Pa*-3 space group was obtained for all investigated samples with varying $Mn^{2+}$ concentrations, both in the $Ce^{3+}$/$Mn^{2+}$ co-doped samples and in the samples doped exclusively with $Ce^{3+}$ (Figure S1)(*36–43*). There are six distinct crystallographic sites occupied by $Ca^{2+}$ ions in $Ca_3Al_2O_6$, as shown in Figure 2b. The Ca(1), Ca(2) and Ca(3) sites are each coordinated by six oxygen atoms, forming octahedral coordination environments. In contrast, the Ca(4), Ca(5), and Ca(6) sites are coordinated by nine, eight and seven oxygen atoms, respectively(*44*, *45*). The numbers of Ca(1)-Ca(6) sites per unit cell are 4, 4, 8, 8, 24 and 24, respectively. Owing to its larger coordination number, the Ca(4) site is expected to be the most favorable site for $Ce^{3+}$ incorporation. However, the molar ratio of the Ca(4), Ca(5), and Ca(6) sites is 1:3:3 within the $Ca_3Al_2O_6$ unit cell, suggesting that the $Ce^{3+}$ ions have a greater probability of occupying the more abundant Ca(5) and Ca(6) sites.

The Raman spectrum recorded at 83 K for the sample co-doped with 5% $Mn^{2+}$ and 5% $Ce^{3+}$ exhibits four main Raman modes in the range of 300-950 $cm^{-1}$ (Figure 2c). The bands below 400 $cm^{-1}$ are mainly attributed to lattice vibrations and external modes involving Ca-O coordination environments. In contrast, the three intense bands centered at approximately 530, 640 and 780 $cm^{-1}$ are associated with internal vibrations of the $AlO_4^{5-}$ units. The band at approximately 530 $cm^{-1}$ can be assigned to vibrations involving Al-O-Al linkages, whereas the feature near 640 $cm^{-1}$ is related to Al-O vibrations of mixed stretching and bending character. The high-frequency band around 780 $cm^{-1}$ is mainly attributed to symmetric Al-O stretching vibrations. The observed splitting and broadening of the Raman bands can be related to the structural complexity of $Ca_3Al_2O_6$, which contains six crystallographically inequivalent Ca sites and two distinct Al sites, resulting in a distribution of metal-oxygen bond lengths and local coordination environments.

SEM images taken for the same sample (Figure 2d) reveal that the material studied consists of irregularly shaped microcrystals characteristic of products obtained by high-temperature solid-state synthesis, without a well-defined or uniform morphology. Furthermore, EDS elemental maps of Ca, Al, O, Mn and Ce confirm a homogeneous spatial distribution of all constituent elements throughout the analyzed region.

To develop a visual luminescent thermometer, it is necessary to obtain distinct luminescence contributions in different spectral ranges. $Ce^{3+}$ ions are among the most well-known luminescent activators exhibiting intense emission in the blue spectral region. The analysis of the spectroscopic properties of $Ca_3Al_2O_6$: $Ce^{3+}$ directly reveals the presence of more than one crystallographic site occupied by $Ce^{3+}$ ions. As shown in Figure 2e, both the excitation and emission spectra strongly depend on the local coordination environment of the $Ce^{3+}$ ions. Under excitation at 460 nm, the material studied exhibits a broad emission band centered at 562 nm, corresponding to the allowed $Ce^{3+}$ $5d^1 \rightarrow 4f^1$ electronic transition(*46–49*). A distinct shoulder appears on the longer wavelength side of the asymmetric emission spectrum. As shown in Figure S2, deconvolution of their emission spectra into Gaussian bands reveals that the spectrum consists of two emission components, which are attributed to the transitions from the 5*d* excited state to the two ground levels of $Ce^{3+}$, i.e. $^2F_{7/2}$ and $^2F_{5/2}$. The energy separation between these peaks (centered at 15,527 $cm^{-1}$ and 17,979 $cm^{-1}$) reaches approximately 2,450 $cm^{-1}$. This value is close to the spin-orbit splitting between the $^2F_{7/2}$ and $^2F_{5/2}$ states usually observed for $Ce^{3+}$ doped phosphors (approximately 2,000 $cm^{-1}$)(*50*), further confirming the assignment of the two emission bands. When monitored at 562 nm, the excitation spectrum exhibits a broad band centered at approximately 460 nm, along with two additional bands at approximately 305 nm and 360 nm. These bands are attributed to the $4f^1 \rightarrow 5d^1$ electronic transitions of $Ce^{3+}$. Upon excitation at 305 nm, the emission spectrum displays several broad emission bands. Gaussian deconvolution reveals that the spectrum consists of four emission components centered at approximately 21,021 $cm^{-1}$, 23,184 $cm^{-1}$, 24,267 $cm^{-1}$, and 26,061 $cm^{-1}$, respectively. These bands can be readily paired, and the corresponding energy differences between them are again, as expected close to the spin-orbit splitting between the $^2F_{7/2}$ and $^2F_{5/2}$ state levels of $Ce^{3+}$ (2163 $cm^{-1}$ between 21021 $cm^{-1}$ and 23184 $cm^{-1}$), and 1794 $cm^{-1}$ (between 24267 $cm^{-1}$ and 26061 $cm^{-1}$). Furthermore, when monitored at 360 nm, the emission spectrum exhibits a broad asymmetric band centered at approximately 505 nm. Analogously in this case Gaussian deconvolution of emission spectra reveals the presence of bands at 19889 $cm^{-1}$, 22672 $cm^{-1}$, 23644 $cm^{-1}$, and 25045 $cm^{-1}$, respectively. These luminescence properties suggest that $Ce^{3+}$ ions occupy multiple crystallographically distinct sites in $Ca_3Al_2O_6$, giving rise to different optical response. The crystal structure of $Ca_3Al_2O_6$ contains six different crystalline sites of $Ca^{2+}$ ions, all of which are coordinated exclusively by oxygen atoms. Among them, Ca(1), Ca(2), and Ca(3) are each six-coordinated, forming octahedral environments, whereas Ca(4), Ca(5), and Ca(6) are coordinated by nine, eight, and seven oxygen atoms, respectively. Furthermore, the numbers of Ca(1), Ca(2), Ca(3), Ca(4), Ca(5), and Ca(6) sites per unit cell are 4, 4, 8, 8, 24, and 24, respectively. Therefore, $Ce^{3+}$ ions are more likely to occupy Ca(5) and Ca(6) sites. This preferential occupation is consistent with the observed multiple emission bands and excitation-dependent luminescence behavior. According to the report of Van Uitert, the positions of the lower d-band edge for $Ce^{3+}$ ion were calculated using the following empirical formula(*51*):

$$E = Q\left[1-\left(\frac{V}{4}\right)^{\frac{1}{V}} 10^{-\frac{near}{80}}\right] \quad (3)$$

where the position in energy ($Q$) is 50,000 cm$^{-1}$ for the lower d-band edge for free $Ce^{3+}$ ion, the valence ($V$) is +3 for the activator $Ce^{3+}$, $n$ is the coordination number of $Ce^{3+}$, the electron affinity ($ea$) of the atoms that form anions is 1.60, and $r$ is the radius of the host cation replaced by the $Ce^{3+}$ ion. The values of $r$ for the $MO_9$, $MO_8$, and $MO_7$ coordination polyhedral are determined by subtracting the ionic radius of $O^{2-}$ from the corresponding average M-O bond lengths. Accordingly, the $E$ values of $Ce^{3+}$ occupying the Ca(4), Ca(5), and Ca(6) sites are calculated to be 23,427 cm$^{-1}$, 21,072 cm$^{-1}$, and 18,397 cm$^{-1}$, respectively. These calculated values indicate that the emission bands observed under 305 nm and 360 nm excitation are attributable to $Ce^{3+}$ ions occupying the Ca(4) and Ca(5) sites, respectively. In contrast, the emission observed under 460 nm excitation is assigned to $Ce^{3+}$ occupying the Ca(6) site (Figure 2f). The emission spectral profiles corresponding to the Ca(4) and Ca(5) sites exhibit concentration-dependent variations with increasing $Ce^{3+}$ concentration (Figure S3). Under 305 nm excitation, the emission from the Ca(4) site dominates at low $Ce^{3+}$ concentrations, whereas the emission profile is reversed at high concentrations, with the Ca(5) site becoming the dominant emission center. Conversely, under 360 nm excitation, the emission band is dominated by the Ca(5) site at low concentration, however, with increasing $Ce^{3+}$ concentration, the Ca(4) site becomes the predominant emission center. This phenomenon might be related to variations in the emission probabilities under different excitation wavelengths. Several factors could contribute to this behavior, including the splitting of energy levels, the degree of distortion of the $Ca^{2+}$ sites, the $Ce^{3+}$-$O^{2-}$ bond lengths, differences in site occupancy, and the distances between $Ce^{3+}$ ions at different $Ce^{3+}$ concentrations. In contrast, the emission spectra recorded under 460 nm excitation show no noticeable changes with varying $Ce^{3+}$ concentrations (Figure S3).

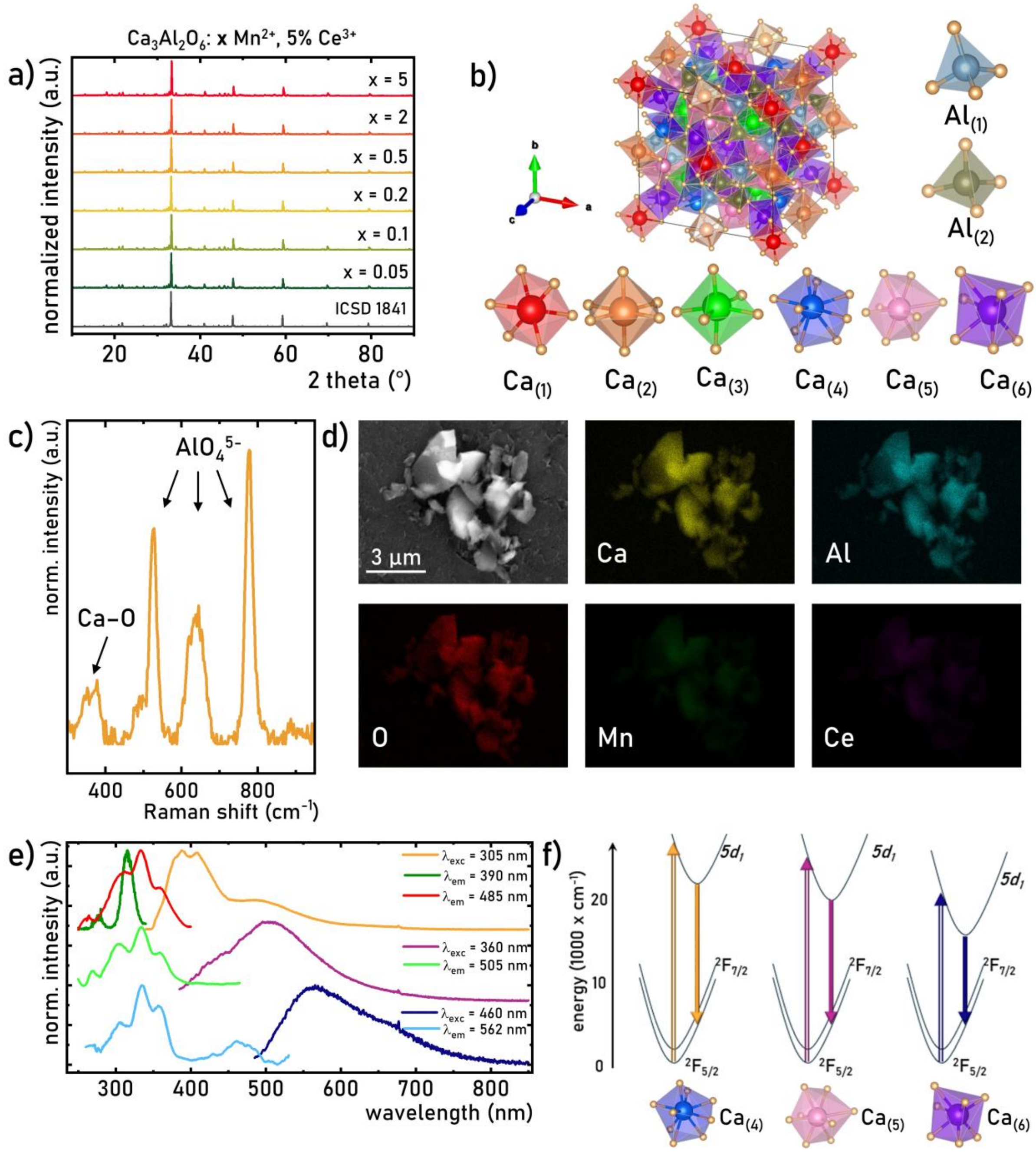

**Figure 2**. Powder XRD patterns of $Ca_3Al_2O_6$:x% $Mn^{2+}$-5% $Ce^{3+}$ with different $Mn^{2+}$ concentrations– a); Visualization of the $Ca_3Al_2O_6$ structure – b); Raman spectrum measured at 83 K – c) and representative SEM image and elemental distributions of Ca, Al, O, Mn and Ce – d) for $Ca_3Al_2O_6$:5% $Mn^{2+}$-5% $Ce^{3+}$; excitation and emission spectra of $Ca_3Al_2O_6$:0.1%$Ce^{3+}$ measured at 93 K at different emission and excitation wavelengths – e); simplified configurational coordination diagram of $Ce^{3+}$ ions localized at different crystallographic sites– f).

**Luminescent properties of $Ca_3Al_2O_6:Ce^{3+}$ and $Ca_3Al_2O_6:Ce^{3+}$, $Mn^{2+}$**

To achieve a temperature-dependent variation in the emission color of $Ca_3Al_2O_6$, which is essential from the perspective of visual luminescence thermometry, $Ca_3Al_2O_6:Ce^{3+}$ was co-doped with $Mn^{2+}$ ions. This approach was motivated by the fact that $Mn^{2+}$ ions located in an octahedral coordination environment exhibit intense red emission associated with the $^4T_1(^4G) \rightarrow {}^6A_1(^6S)$ electronic transition(*52*). According to the Tanabe-Sugano diagram for transition-metal ions with a $3d^5$ electronic configuration, the energy of the $^4T_1$ excited state is strongly dependent on the crystal field strength experienced by $Mn^{2+}$ ions (Figure 3a)(*26*, *27*, *52–55*). In $Ca_3Al_2O_6:Ce^{3+}$, $Mn^{2+}$, the $^4T_1(^4G) \rightarrow {}^6A_1(^6S)$ transition gives rise to a broad emission band centered at around 650 nm, providing a spectrally well-separated emission component with respect to the $Ce^{3+}$ luminescence.

The excitation and emission spectra of the 0.2% $Mn^{2+}$, 5% $Ce^{3+}$ co-doped phosphor (blue and purple curves) and the 5% $Ce^{3+}$ phosphor (orange curves), recorded under different excitation conditions, provide important insight into the underlying excitation and energy-transfer processes (Figure 3b). Upon excitation at 305 nm, the co-doped sample exhibits the characteristic emission bands of $Ce^{3+}$ and $Mn^{2+}$, centered at approximately 485 nm and 650 nm, respectively. These bands are assigned to the $Ce^{3+}$ $5d^1 \rightarrow 4f^1$ and $Mn^{2+}$ $^4T_1(^4G) \rightarrow {}^6A_1(^6S)$ electronic transitions, respectively. The excitation spectrum monitored at the $Ce^{3+}$ emission wavelength of 485 nm consists of several bands in the 250-400 nm spectral range, which can be assigned to the $Ce^{3+}$ $4f^1 \rightarrow 5d$ transitions. In contrast, the excitation spectrum monitored at the $Mn^{2+}$ emission wavelength of 650 nm contains not only the characteristic $Mn^{2+}$ $^6A_1(^6S) \rightarrow {}^4T_2(^4G)$ excitation band but also the $Ce^{3+}$ $4f^1 \rightarrow 5d$ excitation bands. Importantly, the spectral positions of these bands coincide with those observed in the excitation spectrum monitored at the $Ce^{3+}$ emission. The presence of the $Ce^{3+}$ excitation features in the excitation spectrum of $Mn^{2+}$ provides direct evidence for $Ce^{3+} \rightarrow Mn^{2+}$ energy transfer.

The comparison of the emission spectra of $Ca_3Al_2O_6:Ce^{3+}:Mn^{2+}$ recorded at 93 K for phosphors with $Mn^{2+}$ concentrations ranging from 0.05% to 5% indicate that at low $Mn^{2+}$ concentrations, the emission spectra are dominated by the $Ce^{3+}$ luminescence (Figure 3c). However, increasing the $Mn^{2+}$ concentration results in a pronounced enhancement of the $Mn^{2+}$ emission relative to the $Ce^{3+}$. This effect can be explained by the co-existence of two mechanism: (i) growing number of $Mn^{2+}$ emitting centers and (ii) more efficient $Ce^{3+}\rightarrow Mn^{2+}$ energy transfer. The substantial modification of the relative contributions of $Ce^{3+}$ and $Mn^{2+}$ luminescence results in a pronounced change in the overall emission color, as illustrated in Figure 3d. This effect originates from the large spectral separation between the blue $Ce^{3+}$ and red $Mn^{2+}$ emission bands. Consequently, at low $Mn^{2+}$ concentrations the overall emission is dominated by the blue $Ce^{3+}$ component, whereas increasing the $Mn^{2+}$ concentration progressively enhances the red contribution and shifts the resulting emission color toward the red spectral region. Such concentration-controlled tuning of the relative $Ce^{3+}$ and $Mn^{2+}$ emission contributions provides a suitable basis for generating a pronounced temperature-dependent color response and, consequently, for the development of visual luminescence thermometry. This evolution is quantitatively reflected by the $Mn^{2+}$-to-$Ce^{3+}$ integrated emission intensity ratio, which increases systematically with increasing $Mn^{2+}$ concentration (Figure 3e). Specifically, increasing the $Mn^{2+}$ concentration from 0.05% to 5% enhances the $Mn^{2+}$-to-$Ce^{3+}$ emission intensity ratio from 0.16 to 6.4, corresponding to a 40-fold increase.

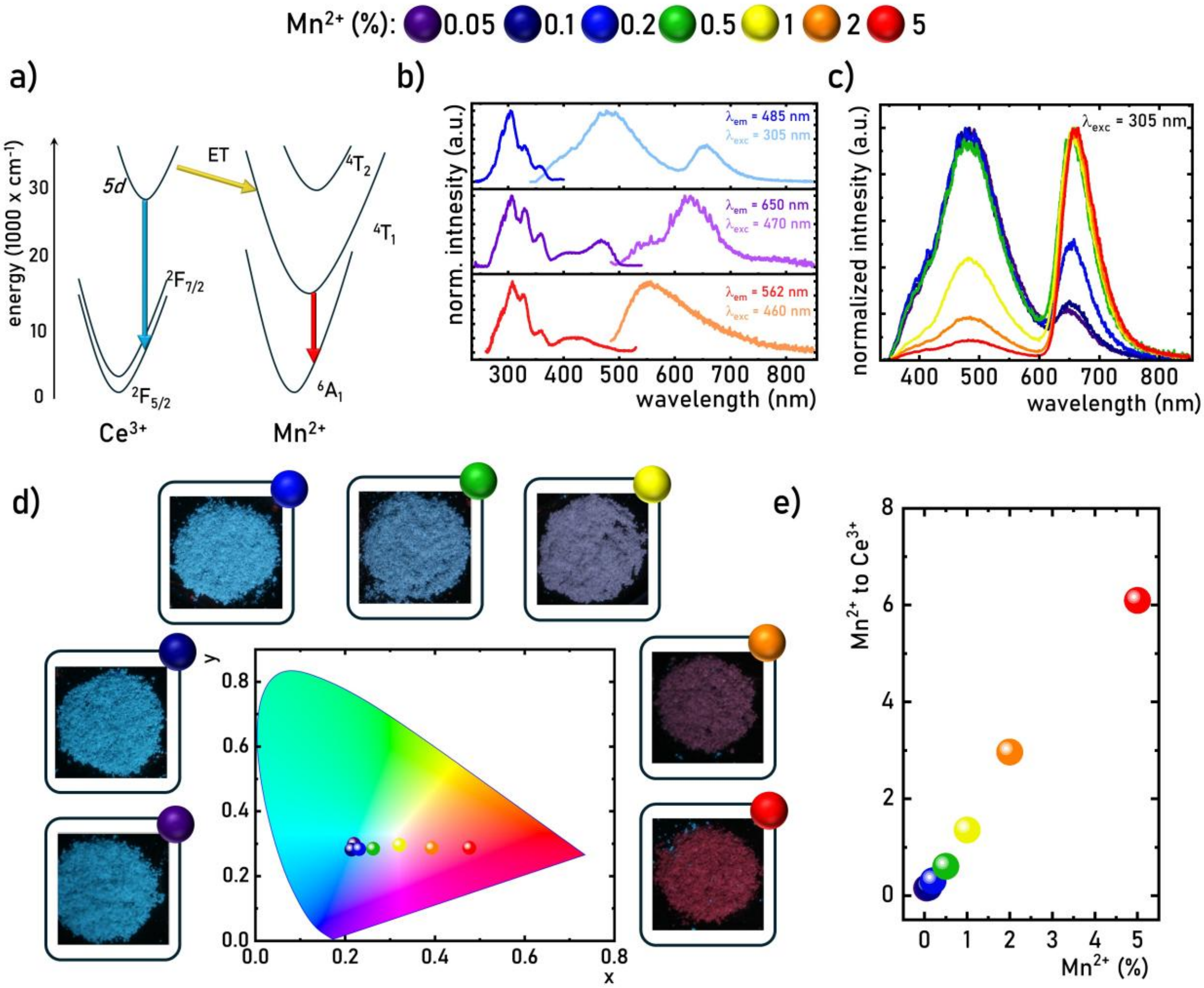

**Figure 3.** Simplified configurational coordination diagram of $Ce^{3+}$ and $Mn^{2+}$ ions – a); excitation and emission spectra of $Ca_3Al_2O_6$:0.2% $Mn^{2+}$ (blue and purple) and $Ca_3Al_2O_6$:5%$Ce^{3+}$ (orange) measured at 93 K at different emission and excitation wavelengths – b); normalized emission spectra– c), CIE1931 chromatic coordinates and corresponding photos of their luminescence -d) and the influence of $Mn^{2+}$ ions concentration on the $Mn^{2+}$ to $Ce^{3+}$ emission intensity ratio for $Ca_{19}Zn_2(PO_4)_{14}$:$Mn^{2+}$, 5%$Ce^{3+}$ with different $Mn^{2+}$ ions concentration measured at 93 K – e).

## Thermometric performance of $Ca_3Al_2O_6$:$Ce^{3+}$, $Mn^{2+}$

To evaluate the potential of the $Ca_3Al_2O_6$:$Mn^{2+}$-$Ce^{3+}$ material for optical thermometry, its temperature-dependent photoluminescence properties were systematically investigated over the *T*-range from 93 to 573 K. As clearly shown in Figure 4a (Figure S4, S5), increasing the temperature leads to a systematic decrease in the emission intensity of both $Ce^{3+}$ and $Mn^{2+}$ ions, although the two luminescent centers exhibit markedly different quenching rates. The thermal evolution of the normalized emission spectra demonstrates that the luminescence of $Ce^{3+}$ decreases rapidly above approximately 380 K, whereas the $Mn^{2+}$ emission band centered at around 680 nm remains clearly observable at higher temperatures (Figure 4b). A quantitative analysis of the integrated emission intensity of $Ce^{3+}$ for different $Mn^{2+}$ concentrations reveals that the $5d\rightarrow4f$ emission decreases to approximately 30% of its initial intensity at 380 K, followed by a dramatic reduction of more than two orders of magnitude upon further heating to 500 K (Figure 4c). Importantly, the concentration of $Mn^{2+}$ has only a negligible influence on the thermal quenching behavior of $Ce^{3+}$, indicating that the thermal depopulation mechanism of the $Ce^{3+}$ excited state is largely independent of the $Mn^{2+}$ content.

In contrast, the thermal quenching of $Mn^{2+}$ luminescence is considerably less pronounced below 380 K, with the emission intensity decreasing to approximately 40% of its initial value at this temperature (Figure 4d). Above 380 K, however, the rate of thermal quenching increases substantially, similarly to the behavior observed for $Ce^{3+}$. In this case, a clear concentration dependence is evident, as increasing the $Mn^{2+}$ content results in progressively faster thermal quenching of the $Mn^{2+}$ emission. This behavior suggests the presence of an additional thermally activated depopulation pathway of the $^4T_1$ excited state of $Mn^{2+}$, which becomes increasingly important at higher dopant concentrations. The most plausible mechanisms are either the formation of $Mn^{2+}$-related structural defects or thermally assisted $Mn^{2+}\rightarrow Ce^{3+}$ energy transfer.

Considering that $Mn^{2+}$ substitutes $Ca^{2+}$ in the investigated host without introducing charge imbalance, the formation of compensating structural defects is expected to be limited. Furthermore, no persistent luminescence was observed in $Ca_{19}Zn_2(PO_4)_{14}$:$Mn^{2+}$-$Ce^{3+}$ indicating the absence of defect-related trapping centers that could significantly influence the emission process. Consequently, thermally activated $Mn^{2+}\rightarrow Ce^{3+}$ energy transfer appears to be the more plausible explanation for the observed concentration-dependent thermal quenching. Although $Ce^{3+}$ and $Mn^{2+}$ possess fundamentally different energy level schemes and therefore are expected to exhibit different intrinsic thermal responses, both ions display a similar increase in the quenching rate above approximately 380 K. This similarity originates from the dominant $Ce^{3+}$→$Mn^{2+}$ energy transfer responsible for populating the excited state of $Mn^{2+}$, causing the thermal population dynamics of $Mn^{2+}$ to closely follow those of the $Ce^{3+}$ 5*d* excited state.

The thermal quenching of $Ce^{3+}$ luminescence is attributed to the thermal ionization of electrons from the 5*d* excited state into the conduction band. This mechanism is supported by the nearly identical temperature dependence of the integrated $Ce^{3+}$ emission intensity observed for phosphors doped solely with $Ce^{3+}$ and those co-doped with $Mn^{2+}$. By fitting the experimental data with the Arrhenius equation, the $5d^1$ excited state of $Ce^{3+}$ can be estimated to lie approximately ~2,100 $cm^{-1}$ below the conduction band. The different thermal responses of $Ce^{3+}$ and $Mn^{2+}$ emission provide the basis for defining the luminescence intensity ratio $LIR_1$ as follows:

$$LIR_1 = \frac{\mathrm{Ce}^{3+}}{\mathrm{Mn}^{2+}} = \frac{\int_{400\mathrm{nm}}^{560\mathrm{nm}} \left(5d \rightarrow {}^4F_{5/2}, {}^4F_{7/2}\right) d\lambda}{\int_{580\mathrm{nm}}^{715\mathrm{nm}} \left({}^4T_1 \rightarrow {}^6A_1\right) d\lambda} \tag{4}$$

For samples containing low concentrations of $Mn^{2+}$, $LIR_1$ exhibits only minor temperature dependence up to approximately 300 K, after which it decreases rapidly, reaching nearly 10% of its initial value at 500 K (Figure 4e). Increasing the $Mn^{2+}$ concentration introduces an initial increase of $LIR_1$ by approximately 10% up to 340 K, followed by thermal quenching at higher temperatures. This behavior is consistent with the concentration-dependent $Mn^{2+}$→$Ce^{3+}$ energy transfer discussed above. To quantify the thermometric performance of the investigated phosphors, the relative thermal sensitivity $S_{R1}$ was calculated as follows:

$$S_R = \frac{1}{LIR} \frac{\Delta LIR}{\Delta T} \cdot 100\% \tag{5}$$

where $\Delta LIR$ stands for the change of $LIR$ corresponding to the change in temperature by $\Delta T$.

The temperature dependence of $S_{R1}$ is dominated by a single broad maximum located around 400 K (Figure 4f). However, both the temperature corresponding to the maximum sensitivity and the magnitude of $S_{R1}$ are strongly influenced by the $Mn^{2+}$ concentration. Specifically, increasing the $Mn^{2+}$ content from 0.05% to 5% shifts $T@S_{Rmax}$ from 440 to 375 K (Figure 4g). The maximum relative thermal sensitivity ($S_{Rmax}$) reaches 0.9% $K^{-1}$ for the sample containing 0.05% $Mn^{2+}$ and increases to 1.25% $K^{-1}$ for the composition with 0.5% $Mn^{2+}$ (Figure 4h). A further increase in the $Mn^{2+}$ concentration results in a gradual decline of $S_{R1max}$, indicating the existence of an optimal dopant concentration for achieving the best thermometric performance.

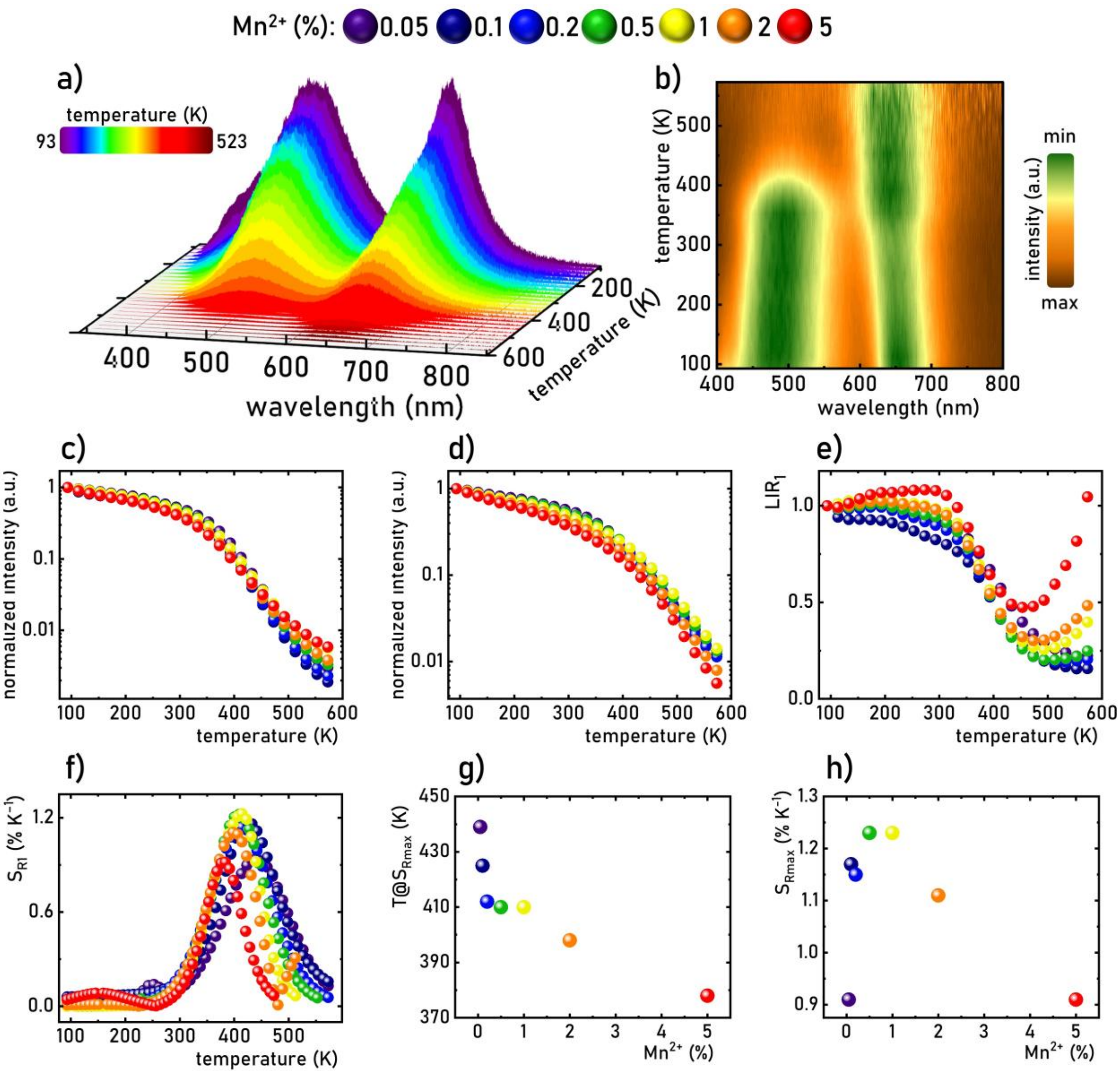

**Figure 4.** Emission spectra of $Ca_3Al_2O_6$:0.5% $Mn^{2+}$, 5% $Ce^{3+}$ measured as a function of temperature – a); thermal map of the normalized emission spectra of $Ca_3Al_2O_6$:0.5% $Mn^{2+}$, 5% $Ce^{3+}$ – b); temperature dependence of integrated emission intensity of $Ce^{3+}$ -c) and of $Mn^{2+}$ ions -d) for the samples with different concentration of $Mn^{2+}$ ions; temperature dependence of $LIR_1$ – e), and corresponding $S_{R1}$ – f); the influence of $Mn^{2+}$ ions concentration on $T@S_{R1max}$ – g) and $S_{R1max}$ – h).

It should be noticed that except thermal quenching of the $Mn^{2+}$ emission intensity observed in $Ca_3Al_2O_6$:0.5% $Mn^{2+}$, $Ce^{3+}$ at elevated temperatures, also a temperature-induced blue shift was recorded (Figure 5a). A similar temperature-induced spectral shift has previously been reported for several $Mn^{2+}$-doped hosts and has been attributed to the thermal population of higher-energy vibronic components of the $^4T_1$ excited state(*15*, *16*, *26*, *52*). This effect becomes particularly evident when the thermal map of the normalized emission spectra is analyzed (Figure 5b). As can be clearly seen, increasing the temperature from 93 to 573 K shifts the spectral position of the $Mn^{2+}$ emission band from 657 nm to 638 nm (Figure S6). Importantly, the magnitude and character of this spectral shift are essentially independent of the $Mn^{2+}$ concentration, supporting its assignment to an intrinsic thermally activated spectroscopic process rather than to concentration-dependent interactions between $Mn^{2+}$ ions.

The temperature-induced shift of the $Mn^{2+}$ emission band can also be exploited for ratiometric luminescence thermometry by defining the luminescence intensity ratio $LIR_2$ using the emission intensities integrated over two selected spectral ranges, which are significantly narrowed compared to the $LIR_1$, namely:

$$LIR_2 = \frac{\mathrm{Ce}^{3+}}{\mathrm{Mn}^{2+}} = \frac{\int_{460\mathrm{nm}}^{505\mathrm{nm}} \left(5d \rightarrow {}^4F_{5/2}, {}^4F_{7/2}\right) d\lambda}{\int_{590\mathrm{nm}}^{635\mathrm{nm}} \left({}^4T_1 \rightarrow {}^6A_1\right) d\lambda} \tag{6}$$

It should be emphasized that these spectral ranges were selected empirically to provide an appropriate compromise between a pronounced temperature dependence of $LIR_2$ and sufficiently high luminescence intensity in both spectral windows. Analysis of the temperature dependence of the luminescence signal integrated over the first spectral range (590-635 nm) reveals that, at low $Mn^{2+}$ concentrations (Figure 5c), the intensity decreases only slightly up to approximately 300 K, whereas above this temperature the thermal quenching rate increases significantly. Importantly, increasing the $Mn^{2+}$ concentration substantially modifies the shape of this dependence. For the sample containing 5% $Mn^{2+}$, the integrated luminescence intensity initially increases with temperature, reaching approximately 125% of its initial value at 253 K, and decreases upon further heating.

In contrast, the luminescence intensity integrated over the second spectral range (650-700 nm) decreases monotonically with increasing temperature and becomes almost completely quenched above 500 K (Figure S8). The distinctly different thermal evolution of the signals collected within the two spectral windows originates predominantly from the temperature-induced blueshift of the $Mn^{2+}$ emission band. As the emission band shifts toward shorter wavelengths, the luminescence signal in the second spectral range decreases more rapidly, whereas the signal collected in the first spectral range becomes thermally stabilized at low $Mn^{2+}$ concentrations and may even increase with temperature at higher $Mn^{2+}$ concentrations. The combination of these opposing thermal responses results in a pronounced temperature dependence of the $LIR_2$ and, consequently, high values of the relative temperature sensitivity - $S_{R2}$.

Importantly, the maximum $S_{R2}$ values are considerably higher than those obtained for the $LIR_1$-based thermometric approach (Figure 5e). The highest $S_{R2}$ value of 1.5% $K^{-1}$ is achieved for the sample containing 0.5% $Mn^{2+}$, demonstrating that the temperature-induced spectral shift of the $Mn^{2+}$ emission band provides a particularly effective ratiometric thermometric readout in the investigated system (Figure 5f).

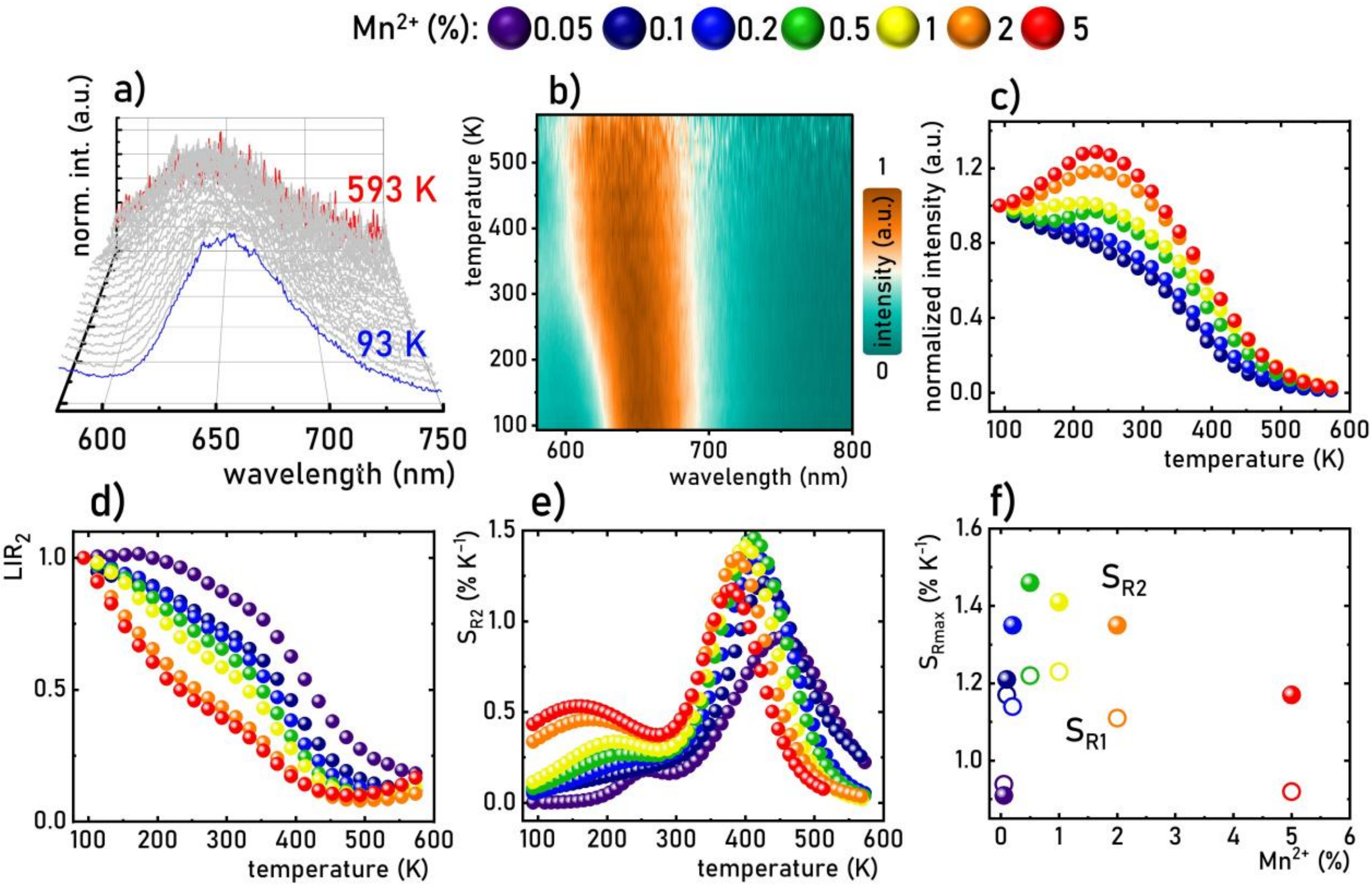

**Figure 5.** Representative thermal evolution of normalized emission band of $Mn^{2+}$ ions– a); thermal map of the emission band of $Mn^{2+}$ emission for $Ca_3Al_2O_6$:0.5% $Mn^{2+}$, 5% $Ce^{3+}$ – b); thermal evolution of integrated emission intensity of $Mn^{2+}$ ions calculated in the 590-635 nm spectral range – c); thermal evolution of $LIR_2$ – d); and corresponding $S_{R2}$ – e) the influence of $Mn^{2+}$ ions concentration on the $S_{R1max}$ and $S_{R2max}$– f).

## Visual thermometry based on $Ca_3Al_2O_6$:$Ce^{3+}$, $Mn^{2+}$

The temperature-dependent thermal quenching of $Ce^{3+}$ and $Mn^{2+}$ emissions induces significant changes in the resulting emission color (Figure 6a). To visualize these temperature-induced color variations, the corresponding CIE1931 chromaticity coordinates were calculated (Figure 6b). The obtained results demonstrate that, at 93 K, all investigated phosphors exhibit gradual redshift in emission color with increasing $Mn^{2+}$ concentration. This shift is associated with the enhanced contribution of $Mn^{2+}$ emission relative to $Ce^{3+}$ emission. Furthermore, increasing temperature causes a systematic redshift of the chromaticity coordinates for all $Mn^{2+}$ concentrations. The effect of $Mn^{2+}$ concentration on the thermal evolution of the *x* and *y* chromaticity coordinates is clearly demonstrated in Figures 6c and S9. Notably, the $Mn^{2+}$ concentration has a more pronounced influence on the $x$ coordinate. With increasing $Mn^{2+}$ concentration, the initial $x$ value at 93 K gradually increases. Compared with the $y$ coordinate, the $x$ coordinate exhibits significantly greater temperature dependence. Although the $y$ coordinate changes only slightly over the entire investigated temperature range, it increases monotonically with increasing temperature. Consequently, the relative sensitivity based on the $x$ chromaticity parameters ($S_{Rx}$) is higher than that based on the $y$ coordinate ($S_{Ry}$) for all investigated samples. The highest relative sensitivities of the $x$ and $y$ coordinates are obtained for the samples co-doped with 0.5% and 0.2% of $Mn^{2+}$, respectively, reaching maximum values of $S_{Rxmax}$ = 0.45 $\%K^{-1}$ at 410 K and $S_{Rymax}$ = 0.14 $\%K^{-1}$ at 95 K (Figure 6d and S9). Furthermore, the temperatures corresponding to the maximum relative sensitivities of $S_{R1}$, $S_{R2}$ and $S_{Rx}$, exhibit a similar trend, shifting toward lower temperatures with increasing $Mn^{2+}$ concentration. Additionally, it can be clearly seen that at elevated temperature the temperature at which $S_{Rmaxx}$ was observed decrease with dopant concentration (Figure 6f). On the other hand, in the case of the $T@S_{Rmaxy}$ only some fluctuation around 250 K can be observed. Another parameter which describes temperature-induced changes in the emission color is the chromaticity difference (*CD*), defined as follows:

$$CD = \sqrt{(x_f - x_i)^2 + (y_f - y_i)^2} \tag{7}$$

where $x_f$, $y_f$ and $x_i$, $y_i$ represents the final (high temperature) and the initial (low temperature) chromatic coordinates. The *CD* parameter initially increases from 0.15 for the sample with 0.05% of $Mn^{2+}$ to 0.22 for the one with 0.5% of $Mn^{2+}$, and then gradually decreases with further increasing $Mn^{2+}$ concentration (Figure 6g).

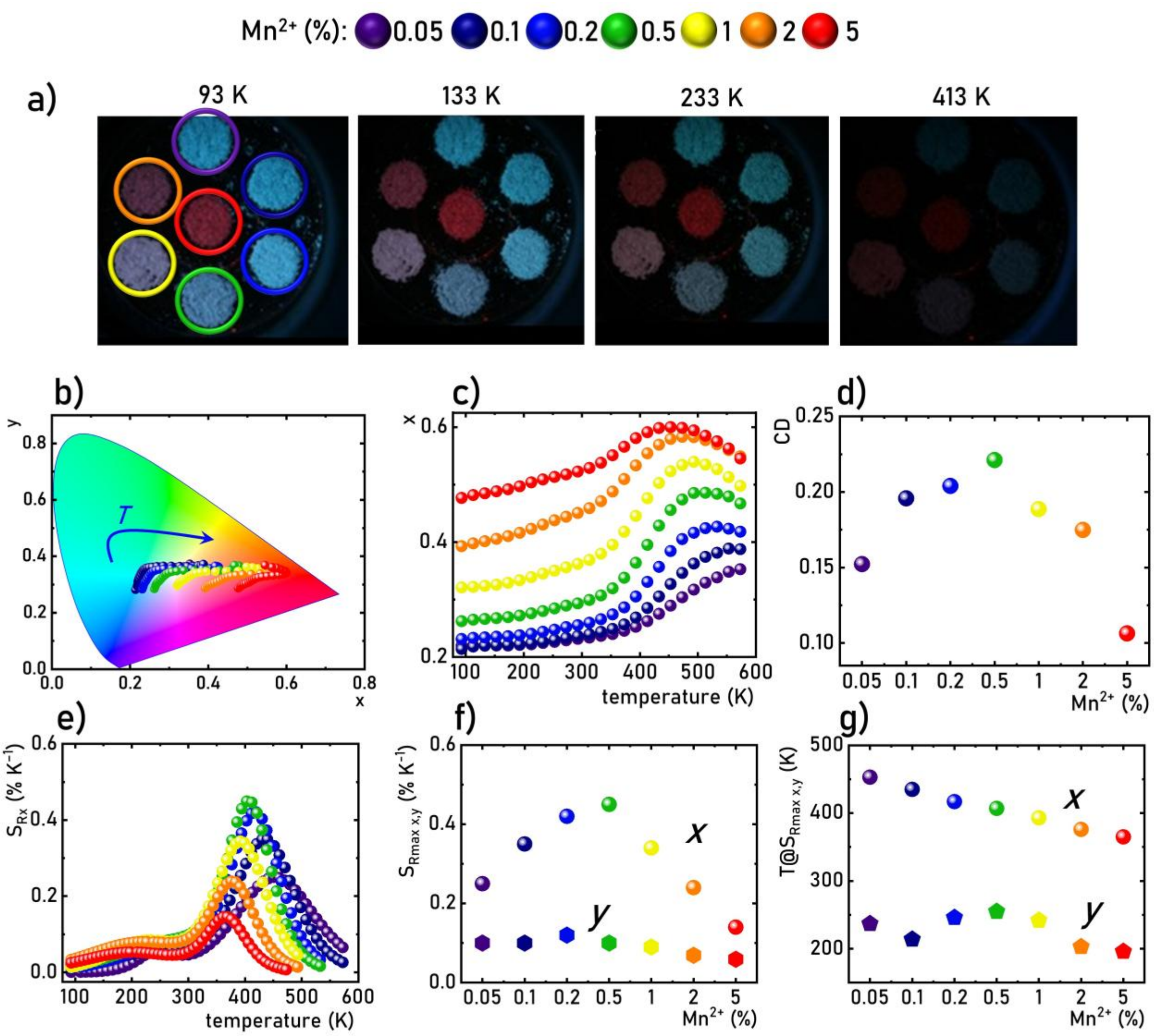

**Figure 6**. Representative photographs of the luminescence of $Ca_3Al_2O_6$ co-doped with x% of $Mn^{2+}$ (x = 0.05-5) and 5% of $Ce^{3+}$ at different temperatures - a); the influence of temperature on the CIE1931 chromatic coordinates of the analyzed phosphors – b); the thermal evolution of *x* chromatic CIE1931 coordinates of $Ca_3Al_2O_6$:x%$Mn^{2+}$, 5%$Ce^{3+}$ for different $Mn^{2+}$ concentrations - c); luminescence colour change (*CD*) as a function of $Mn^{2+}$ concentration - d); thermal evolution of $S_{Rx}$ -e); the influence of the $Mn^{2+}$ ions concentration on the $S_{Rmaxx,y}$ -f) and $T@S_{Rmaxx,y}$ -g).

## Conventional RGB-based thermometry

As demonstrated above, the temperature-induced variation in the emission color of $Ca_3Al_2O_6$ co-doped with x% of $Mn^{2+}$ (x = 0.05-5) and 5% of $Ce^{3+}$ can be quantitatively evaluated through analysis of the CIE 1931 chromaticity coordinates. However, from a practical perspective, this approach may be inconvenient, as determination of the CIE 1931 parameters requires acquisition of the luminescence spectrum of the phosphor. Therefore, the same luminescence thermochromism observed in $Ca_3Al_2O_6:Mn^{2+}$, 5%$Ce^{3+}$ can be exploited more straightforwardly through direct analysis of digital images of the emitted light. During image acquisition, a conventional digital camera inherently records two-dimensional intensity maps in the red ($R$), green ($G$), and blue ($B$) channels. Owing to their distinct spectral response ranges, these channels can selectively probe the contributions of $Ce^{3+}$ and $Mn^{2+}$ luminescence. Consequently, pixel-by-pixel calculation of the ratio between appropriately selected *RGB* intensity maps can provide a simple approach for spatially resolved temperature determination, eliminating the need for spectroscopic detection.

Representative *RGB* intensity maps extracted from images of $Ca_3Al_2O_6:Mn^{2+}$, 5%$Ce^{3+}$ containing different $Mn^{2+}$ concentrations and recorded at 93 K are presented in Figure 7a. The analysis of the temperature evolution of the normalized *RGB* channels pixel intensity for all samples reveals that these intensities decrease at elevated temperatures (Figure 7b, c, Figure S10). However, it can be noticed that $Mn^{2+}$ ions concentration affects these trends. For samples with low $Mn^{2+}$ concentrations, namely, 0.05 - 1% (Figure 7b), the three channels exhibit very similar rates of thermal changes. In contrast, at higher $Mn^{2+}$ concentrations (2-5 % of $Mn^{2+}$), the responses of the individual channels become increasingly distinct, with the $R$ channel showing markedly different temperature dependence compared with the $G$ and $B$ channels. These results indicate that the *RGB* channel intensities are not only temperature-dependent, but can also be tuned through the $Mn^{2+}$ concentration. Looking forward to developing an optical thermometer based on the *RGB* channels, different thermometric parameters derived from these channels were considered. In principle, a ratiometric parameter defined from two *RGB* channels should also exhibit a temperature dependence if the individual channels respond differently to temperature. However, for samples with $Mn^{2+}$ concentrations between 0.05% and 1%, the *R*, *G* and *B* channels display very similar temperature trends. As a result, the corresponding ratiometric parameters remain nearly constant with temperature, limiting their suitability as thermometric parameters. These samples were therefore excluded from the following analysis.

In the case of the high dopant concentration both *B/R* and *G/R* ratiometric parameters reveal monotonic thermal trend up to 333 K (Figure 7d, e). Therefore, they can be exploited for temperature sensing. In order to perform a calibration procedure, a phenomenological straight line calibration curve for both *B/R* and *G/R* was adopted as shown in Figure 7d and e. The temperature dependence of the $S_R$, depicted in Figure 7c, d, presents an increase with temperature, reaching maximum values of 0.818% $K^{-1}$ and 0.371 % $K^{-1}$ at 333 K, for *B/R* and *G/R* respectively (Figure 7c, d). It is worth noting that the relative sensitivity associated with the *B/R* ratio decreases as the $Mn^{2+}$ concentration decreases, with a similar trend observed across all three samples. A similar behaviour was observed for the *G/R* ratio. The only exception was sample with 0.5% of $Mn^{2+}$ (Figure S11) concentration, for which the *B/G* ratio was analysed instead of *G/R*. Once again, these results suggest that the sensitivity of the optical thermometer can be tuned by varying the $Mn^{2+}$ concentration.

**Implementation of MLR to RGB-based luminescence thermometry**

As it can be seen the obtained thermal sensitivities of the ratiometric parameter based on *RGB* analysis is not very high in the case of $Ca_3Al_2O_6:Mn^{2+}, Ce^{3+}$. Therefore, we introduce here the multiple linear regression (MLR) approach to exploit the complementary temperature-dependent information encoded in different *RGB* channel ratios and enhance the thermometric performance beyond that achievable using individual ratiometric parameters. Although, MLR was already used in temperature sensing(*33–35*, *56*), to the best of our knowledge, the combination of multiple ratiometric *RGB* parameters with MLR has not previously been explored for optical thermometry. Considering the nearly linear temperature dependence exhibited by the *RGB* channel ratios, their simultaneous analysis through MLR provides a straightforward multiparametric framework to improve the sensitivity of the system. MLR has been successfully introduced in luminescence thermometry as a means of evaluating the combined contribution of several independent temperature-dependent variables to a single experimental response, boosting as a result the performance of the optical sensor. Within this framework, the MLR model can be expressed as:

$$y = \beta_0 + \sum_{i=1}^{n} \beta_i x_i + \varepsilon \quad (7)$$

where $y$ represents the dependent variable (in this study, temperature), $x_i$ are the independent variables, also referred to as predictors, $\beta_i$ are the regression coefficients associated with each predictor, and ε denotes the residual random error. The regression coefficients quantify the variation in the dependent variable resulting from a one-unit change in the corresponding predictor, while the remaining predictors are kept constant. The error term accounts for unobserved factors influencing the dependent variable that are not explicitly included in the model. In the present work, the predictors correspond to the ratiometric parameters based on the *RGB* channels, while temperature is treated as the response variable. To ensure the validity of the linear regression model, the selected predictors were required to satisfy two main criteria: (i) they must exhibit a linear dependence on temperature, and (ii) they must share the same operational temperature range. In the case of $Ca_3Al_2O_6$:5%$Ce^{3+}$:5% $Mn^{2+}$ two selected thermometric parameters, namely *B/R* and *G/R* exhibit linear dependence with temperature and share the same operational ranges, namely from 93 to 333 K. The combination of these two predictors was carried out following the procedure described in the literature, and the corresponding *β*-coefficients were calculated(*31*, *57*). *B/R* presents a higher *β*-weight than *G/R* (76.1% and 23.9% respectively, Figure 7f). According to the model, *B/R* plays a more significant role in the estimation of the temperature for a given value of *G/R*. As it can be seen the implementation of MLR for this phosphor enables to obtain the good correlations between the measured temperatures and those predicted by the MLR model (Figure 7g). Additionally, it can be clearly shown that the $S_R$ obtained based on MLR are significantly much higher comparring to the one obtained based on sloely *B/R* and *G/R* for all $Mn^{2+}$ ions concentrations (Figure 7h, Figure S12). The MLR model yielded a maximum $S_R$ value of 1.70% $K^{-1}$ for 5% $Mn^{2+}$, representing approximately a threefold improvement compared with the $S_R$ values obtained from single-parametric sensing in the same temperature range (0.63% $K^{-1}$ for *B/R*). Overall, these results demonstrate that the advantages previously reported for MLR-based multiparametric optical sensing can also be extended to *RGB*-based thermometry and the obtained thermal sensitivity is the highest among all up to now reported $Mn^{2+}$ based luminescent thermometers (Table 1). In particular, the simultaneous combination of different *RGB* ratiometric parameters allows the complementary temperature-dependent information contained in the individual colour channels to be exploited, resulting in a substantial improvement of the $S_R$. This work therefore introduces the use of MLR for the combined analysis of *RGB* ratiometric parameters in optical thermometry, providing a new strategy to improve the performance of *RGB*-based luminescent thermometers beyond that achievable using individual channel ratios.

**Table 1** Comparison of the thermometric performance of RGB luminescence thermometry.

| Phosphors | Channels | $S_R$ [% $K^{-1}$] | Ref |
|---|---|---|---|
| **$La_2MoO_6$:$Yb^{3+}$/$Tm^{3+}$** | R/G | 1.8 | (58) |
| **$Ca_{19}Ce(PO_4)_{14}$:$Mn^{2+}$, $Ce^{3+}$** | R/G | 0.69 | (*15*) |
| **$Ca_{19}Zn_2(PO_4)_{14}$:$Mn^{2+}$, $Ce^{3+}$** | R/G | 0.72 | (59) |
| **$Ca_3Al_2O_6$: $Mn^{2+}$, $Ce^{3+}$** | B/R | 0.8 | This work |
| **$Ca_3Al_2O_6$: $Mn^{2+}$, $Ce^{3+}$** | G/R | 0.4 | This work |
| **$Ca_3Al_2O_6$: $Mn^{2+}$, $Ce^{3+}$** | MLR | 1.7 | This work |

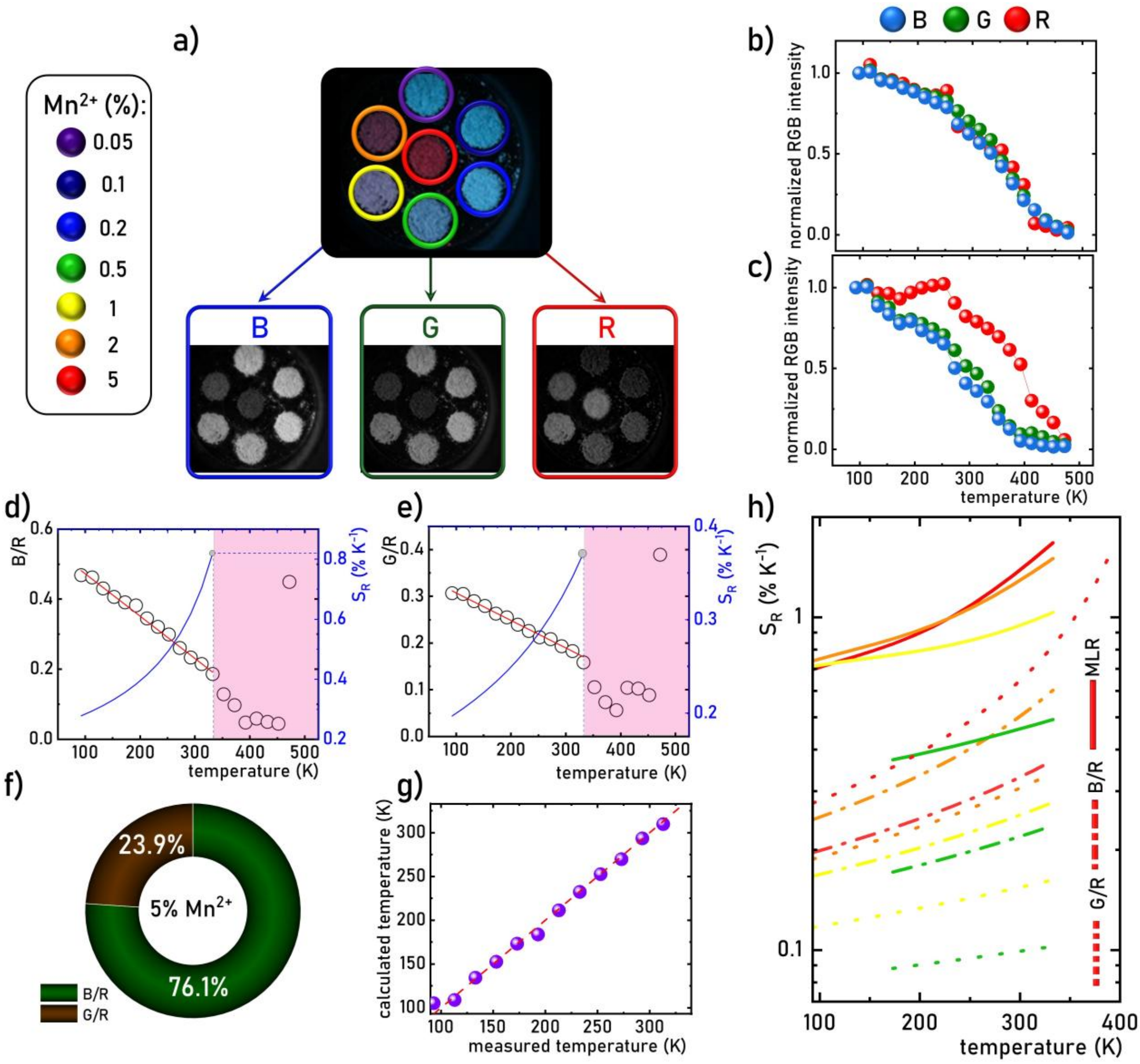

**Figure 7**. Representative photographs of luminescence of the $Ca_3AlO_2O_6$:$Mn^{2+}$, 5% $Ce^{3+}$ samples taken at 93 K and the extraction of the intensity recorded in the R, G, B channels -a); temperature dependance of the normalized RGB channels pixel intensity for the $Ca_3AlO_2O_6$:0.05% $Mn^{2+}$, 5%$Ce^{3+}$ - b) and $Ca_3AlO_2O_6$:5% $Mn^{2+}$, 5% $Ce^{3+}$ materials -c); temperature dependance of the *B/R* – d) and *G/R* – e) ratiometric parameters (open symbols) and their corresponding relative thermal sensitivity, $S_R$ (blue solid line) of $Ca_3Al_2O_6$:5% $Mn^{2+}$, 5% $Ce^{3+}$ sample (red solid lines in both stand for the linear fits as the best fits to the experimental data, see Table S1. The pale pink shadowed region represents the excluded temperature region.); doughnut chart of the relative weight (values on the chart are the $\beta$-weights) of the thermometric parameters *B/R* and *G/R* considered for the MLR analysis within the range of 93-333 K for the sample $Ca_3Al_2O_6$:5% $Mn^{2+}$, 5% $Ce^{3+}$ -f); correlation between the measured temperature and the calculated temperature by the MLR model – g); the influence of the dopant concentration on the thermal dependence of $S_R$ -h).

## DISCUSSION

In this work, we introduce, for the first time, a multiple-linear-regression-enhanced RGB-based luminescence thermometry strategy that combines multispectral image analysis with 2D spatial temperature mapping using a conventional digital camera. We used a $Ca_3AlO_2O_6$ material co-doped with $Mn^{2+}$ and $Ce^{3+}$ ions as a representative system, in order to systematically investigate the effect of dopant concentration and temperature on its thermometric performance for visual thermal sensing and imaging. The incorporation of $Ce^{3+}$ and $Mn^{2+}$ ions into the investigated system enables luminescence in the blue and red spectral ranges, respectively, originating from the $5d^1 \rightarrow 4f^1$ transition of $Ce^{3+}$ and the $^4T_1(^4G) \rightarrow {}^6A_1(^6S)$ transition of $Mn^{2+}$. Although the spectroscopic analysis of $Ca_3AlO_2O_6:Ce^{3+}$ revealed the presence of several crystallographic sites occupied by $Ce^{3+}$ ions, the dominant luminescence contribution was found to originate from $Ce^{3+}$ ions located at the Ca(5) site.

Comparison of the excitation spectra of $Ca_3AlO_2O_6:Ce^{3+}$ and $Ca_3AlO_2O_6:Ce^{3+},Mn^{2+}$ revealed that $Ce^{3+} \rightarrow Mn^{2+}$ energy transfer constitutes the dominant pathway responsible for populating the $^4T_1(^4G)$ excited state of $Mn^{2+}$ ions. Consequently, owing to the increasing number of $Mn^{2+}$ luminescent centers and the efficient $Ce^{3+} \rightarrow Mn^{2+}$ energy transfer, increasing the $Mn^{2+}$ concentration progressively enhances the intensity of the $Mn^{2+}$-related $^4T_1(^4G) \rightarrow {}^6A_1(^6S)$ emission band relative to the $Ce^{3+}$ emission. As a result, the emission color of $Ca_3AlO_2O_6:Ce^{3+},Mn^{2+}$ can be continuously tuned from blue to red by increasing the $Mn^{2+}$ concentration.

Temperature-dependent spectroscopic analysis revealed that $Mn^{2+}$ luminescence exhibits higher thermal stability than the $Ce^{3+}$ emission. At the same time, the comparable temperature dependence of the emission intensities of both ions observed up to approximately 400 K can be explained by thermalization of the $Ce^{3+}$ excited state via the conduction band. The different thermal responses of $Ce^{3+}$ and $Mn^{2+}$ luminescence enable ratiometric temperature readout based on the intensity ratio of their respective emission bands ($LIR_1$), yielding a maximum relative sensitivity of $S_{Rmax}$ = 1.24 % $K^{-1}$ for the sample containing 0.5% $Mn^{2+}$. Importantly, increasing the $Mn^{2+}$ concentration results in a monotonic shift of the temperature corresponding to the maximum relative sensitivity, $T@S_{Rmax}$, from 442 K for 0.05% $Mn^{2+}$ to 378 K for 5% $Mn^{2+}$, demonstrating the possibility of tailoring the optimal operating temperature range through appropriate control of the dopant concentration.

In addition to the changes in emission intensity, increasing temperature induces a pronounced blueshift of the $Mn^{2+}$ emission band, which can be attributed to the thermal population of higher-energy vibronic components of the $Mn^{2+}$ excited state. This effect was exploited to define an additional thermometric parameter, $LIR_2$, in which only the short-wavelength portion of the $Mn^{2+}$ emission band was considered. This approach significantly enhanced the thermometric response compared with that obtained using $LIR_1$, resulting in $S_{Rmax}$ = 1.5 % $K^{-1}$ for the sample containing 0.5% $Mn^{2+}$. The pronounced temperature-induced modification of the emission color of $Ca_3AlO_2O_6:Ce^{3+},Mn^{2+}$ additionally enables temperature determination through analysis of the CIE 1931 chromaticity coordinates, with maximum relative sensitivities of $S_{Rx,max}$ = 0.45% $K^{-1}$ *and* $S_{Ry,max}$ = 0.14% $K^{-1}$.

Notably, the thermochromic response of the $Ca_3AlO_2O_6:Ce^{3+}$, $Mn^{2+}$ material can be directly analyzed using a conventional digital camera, which is crucial from the perspective of practical thermal imaging. In this approach, ratios of the two-dimensional intensity maps recorded by the *R*, *G*, and *B* channels provide spatially resolved information on the temperature distribution and therefore enable straightforward *RGB*-based thermal imaging without the need for spectrally resolved detection. Furthermore, for the first time, a multiple linear regression (MLR) approach was implemented to enhance the thermometric performance of *RGB*-based thermal imaging. By simultaneously exploiting the temperature-dependent information encoded in the *RGB* channels, the MLR-based approach resulted in an approximately 3-fold enhancement of the relative sensitivity, reaching $S_{Rmax}$ = 1.7 % $K^{-1}$ for the sample containing 5% $Mn^{2+}$. Overall, these results demonstrate that the combination of luminescence thermochromism, *RGB*-based image analysis, and MLR provides a simple yet highly sensitive strategy for spatially resolved thermal sensing and imaging using a conventional digital camera.

## MATERIALS AND METHODS

### Synthesis

The phosphor materials, $Ca_3Al_2O_6$ co-doped with $Mn^{2+}$ and $Ce^{3+}$, were synthesized by a conventional solid-state reaction. CaO (Thermo Fisher Scientific, 99.998%), $Al_2O_3$ (Thermo Fisher Scientific, 99.995%),), $MnCO_3$ (Thermo Fisher Scientific, 99.985%) and $Ce(NO_3)_3 \cdot 6H_2O$ (Alfa Aesar, 99.998%) were used as starting materials. According to the designed compositions, the proper amounts of these above powders were weighed and ground in an agate mortar. The grounded mixtures were transferred to corundum crucibles and annealed in an air atmosphere at 1573 K for 4 hours, after that in a reducing atmosphere at 1573 K for 4 hours. The final samples were naturally cooled to the room temperature and grounded to a powder for further evaluations.

### Characterization

The powder X-ray diffraction (XRD) patterns were recorded using a PANalytical X'Pert Pro diffractometer using Ni-filtered Cu Kα radiation (V = 40 kV, I = 30 mA). Measurements were performed in the 2θ = 10 - 90º range with a 30 min measurement time. Scanning electron microscopy (SEM) was taken by FEI Nova NanoSEM 230 equipped with an EDAX Genesis XM4 energy dispersive spectrometer to verify the morphology of the samples and the distribution of its elements by Energy-Dispersive X-ray Spectroscopy (EDS) (V = 30 kV for SEM and V = 5 kV for EDS mapping). Samples were prepared by dispersing the powder in a few drops of methanol. A drop of the resulting suspension was deposited onto a carbon stub and allowed to dry. This preparation method ensured stable positioning of the sample during SEM imaging while minimizing charging effects under electron beam irradiation.

The optical properties including emission, excitation spectra and luminescence kinetics were measured by a FLS1000 Fluorescence Spectrometer from Edinburgh Instruments, equipped with a 450 W xenon lamp and μFlash pulsed lamp as excitation sources, and an R928 photomultiplier tube from Hamamatsu as a detector. The temperature-dependent measurements were performed using a THMS 600 heating-cooling stage from Linkam, providing temperature stability of 0.1 K and a set point resolution of 0.1 K.

The digital images were taken using a Canon EOS 400D camera using a 10 s integration time, 14.3 lp/nm spatial resolution. After acquiring the luminescence images, the individual R, G and B channels were extracted and emission maps based on their intensity ratios were subsequently generated. All image processing steps were performed using IrfanView 64 (version 4.51).

**Acknowledgements**

This work was supported by the Foundation for Polish Science under First Team FENG.02.02-IP.05-0018/23 project with funds from the 2nd Priority of the Program European Funds for Modern Economy 2021-2027 (FENG). Authors would like to acknodledge dr Damian Szymanski for help in the EDX analysis.